\documentclass{PoS}
\usepackage{amsfonts,amssymb,amsmath,bm}
\usepackage{graphicx}

\newcommand{\be}{\begin{equation}}
\newcommand{\bea}{\begin{eqnarray}}
\newcommand{\ee}{\end{equation}}
\newcommand{\eea}{\end{eqnarray}}
\newcommand{\bpi}{\begin{picture}}
\newcommand{\bce}{\begin{center}}
\newcommand{\epi}{\end{picture}}
\newcommand{\ece}{\end{center}}

\def\gtree{\Gamma^{(0)}}
\def\gtreeb{\widetilde{\Gamma}^{(0)}}
\def\gfullb{\widetilde{\Gamma}}
\def\bqq{\widetilde{\bm{\Gamma}}}

\def\bcj{J}
\def\bcjb{\widetilde{\bcj}}

\def\gb{\bm{\Gamma}}

\title{A gauge-technique Ansatz for the three gluon vertex of the background field method}

\ShortTitle{Ansatz for the three gluon vertex}

\author{\speaker{Joannis Papavassiliou}
\\
Department of Theoretical Physics and IFIC,\\
University of Valencia,\\ 
E-46100, Valencia, Spain.
\\
        E-mail: \email{Joannis.Papavassiliou@uv.es}}

\abstract{
The vertex connecting one background gluon with two quantum ones
constitutes a central 
ingredient in the gauge-invariant Schwinger-Dyson equation that determines the non-perturbative 
dynamics of the gluon propagator. This vertex satisfies a Ward identity with 
respect to the background gluon, and a Slavnov-Taylor identity with respect 
to the two quantum gluons.  We present a complete Ansatz for this 
vertex, which satisfies  both aforementioned identities. 
This entire construction depends crucially 
on a set of constraints relating the various form-factors of the ghost Green's functions  
appearing in the Slavnov-Taylor identity satisfied by the vertex. The validity of these  
constraints is demonstrated to all orders.}

\FullConference{The many faces of QCD\\
		November 2-5, 2010\\
		Gent Belgium}

\begin{document}

\maketitle

\section{Introduction}

\begin{figure}[!t]
\begin{center}
\includegraphics[scale=.6]{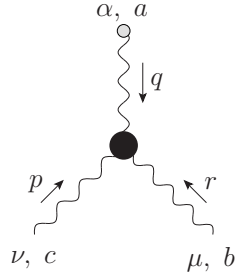}
\end{center}
\caption{\label{3g-vertex}The $BQQ$ three-gluon vertex. The background leg is indicated by the gray circle.}
\end{figure}

The three-gluon vertex describing the interaction of one background ($B$)  
and two quantum ($Q$) gluons (``$BQQ$ vertex'', for short) [see Fig.~\ref{3g-vertex}]
is of particular interest, because it constitutes a key ingredient 
for understanding certain important aspects of non-perturbative QCD. 

As has been explained in a series of recent articles~\cite{Aguilar:2006gr}, the fully dressed version of the $BQQ$ vertex 
is instrumental for the gauge-invariant truncation
the Schwinger-Dyson equations (SDE) obtained within the general formalism based on the 
Pinch Technique (PT)~\cite{Cornwall:1981zr} and the Background Field Method (BFM)~\cite{Abbott:1980hw},  
and especially for the  
crucial transversality properties displayed by the  SDE governing the gluon self-energy.
In particular, and contrary to what happens in the conventional formulation, 
the ``one-loop dressed'' subset 
of (only gluonic!) diagrams, 
(corresponding to the first step in the aforementioned SDE truncation),
furnishes an exactly transverse  gluon self-energy.

The SDEs of the PT-BFM  
have been particularly successful in reproducing recent lattice data~\cite{Aguilar:2008xm,RodriguezQuintero:2010ss}, 
which clearly indicate that 
the gluon propagator and the ghost dressing function 
of Yang-Mills in the Landau gauge are infrared finite  
both in $SU(2)$~\cite{Cucchieri:2007md,Bogolubsky:2007ud} and in $SU(3)$. 
The non-perturbative form of the $BQQ$ vertex is 
essential for accomplishing this task. 
Indeed, the way the gluon acquires a dynamically generated (momentum-dependent) mass (first paper in~\cite{Cornwall:1981zr}), 
which, in turn, accounts for the infrared  finiteness of the aforementioned 
Green's functions, is determined by a subtle interplay between various crucial features  
of this special vertex~\cite{Aguilar:2009ke}. 

To be sure, the non-perturbative behavior of the $BQQ$ vertex is determined by its
own SDE equation, which contains the various multiparticle kernels appearing 
in the ``skeleton expansion''. However, for practical purposes, one is 
forced to resort to an Ansatz for this vertex, obtained through the so-called 
``gauge-technique''~\cite{Salam:1963sa}. 

The idea behind the gauge-technique is fairly simple, especially in an 
abelian context: one constructs an expression for the unknown 
vertex out of the ingredients appearing in the Ward identity (WI) it satisfies. 
These ingredients must be put together in a way such that 
the resulting expression satisfies the WI automatically. 
The most typical example of such a construction is found in the case of the three-particle vertex
of scalar QED, describing the interaction of a photon with a pair of charged scalars.
This vertex, to be denoted by $\gb_{\mu}$,  satisfies the abelian all-order WI
\be
q^{\mu}\gb_{\mu}= {\mathcal D}^{-1}(k+q) -{\mathcal D}^{-1}(k) \,, 
\label{sward}
\ee
where ${\mathcal D}(k)$  is the fully-dressed propagator of the scalar field.
Thus, in this case,  the gauge-technique Ansatz for $\Gamma_{\mu}$, obtained by Ball and Chiu \cite{Ball:1980ay}, 
after ``solving'' the above WI, under the additional  
requirement of not introducing kinematic singularities, is 
\be
\gb_{\mu}= \frac{(2k+q)_{\mu}}{(k+q)^2-k^2}\left[{\mathcal D}^{-1}(k+q) -{\mathcal D}^{-1}(k)\right],
\label{strans_vert}
\ee
which clearly satisfies Eq.~(\ref{sward}).

Returning to the case at hand, and according to the philosophy explained above,  
one must construct
the $BQQ$ vertex out of the ingredients appearing  in the 
WI and Slavnov-Taylor identity (STI) it satisfies (third paper in~\cite{Aguilar:2006gr}), and in such a way that  
these identities are automatically satisfied. 
This is a complicated task, because some of the ingredients appearing in the STI 
(i.e., the Green's functions originating from the ghost sector)
are themselves constrained by yet another  set of (largely unexplored) WIs and STIs,  
which must be exactly preserved, or else the entire construction will collapse, 
or major complications will appear in subsequent steps of the SDE treatment.

The purpose of this talk is to describe in some detail how the longitudinal part of the  
$BQQ$ vertex can be obtained by ``solving'' the corresponding WI and STI, and the important 
role played by the constraints relating the various ghost Green's functions appearing 
in the STI. A complete account of the important implications of this construction on the SDE of the 
gluon propagator will be given elsewhere~\cite{Binosi:2011wi}.

\section{The $BQQ$ vertex and its basic properties}

 The $BQQ$ vertex constitutes without a doubt one of the most fundamental ingredients of the PT, making its appearance already at the basic level of the one-loop construction. 
Specifically,  defining the tree-level conventional 
three-gluon vertex through the expression (all momenta entering)
\bea
i\Gamma^{(0)}_{A^a_\alpha A^b_\mu A^c_\nu}(q,r,p)&=&gf^{abc}\gtree_{\alpha\mu\nu}(q,r,p)\nonumber \\
\gtree_{\alpha\mu\nu}(q,r,p)&=&g_{\mu\nu}(r-p)_\alpha  +
g_{\alpha\nu}(p-q)_\mu+g_{\alpha\mu}(q-r)_\nu,
\eea
the diagrammatic rearrangements giving rise to the PT Green's functions (propagators and vertices) stem exclusively from the  characteristic decomposition~\cite{Cornwall:1981zr}
\begin{eqnarray}
\gtree_{\alpha\mu\nu}(q,r,p)&=& \gtreeb_{\alpha\mu\nu}(q,r,p)+ (1/\xi)
\Gamma^{{\rm P}}_{\alpha\mu\nu}(q,r,p),     \nonumber  \\ 
\gtreeb_{\alpha\mu\nu}(q,r,p)&=&g_{\mu\nu}(r-p)_\alpha  +
g_{\alpha\nu}(p-q+ r/\xi)_\mu+g_{\alpha\mu}(q-r- p/\xi)_\nu, \nonumber \\    
\Gamma^{{\rm P}}_{\alpha\mu\nu}(q,r,p)&=& g_{\alpha\mu} p_\nu - g_{\alpha\nu}r_\mu.
\label{deco}
\end{eqnarray}
In the equations above, $\xi$ represents the gauge fixing parameter that  appears also in the definition of the (full) gluon propagator $\Delta^{ab}_{\mu\nu}(q)=\delta^{ab}\Delta_{\mu\nu}(q)$, with
\be
i\Delta_{\mu\nu}(q)=- i\left[P_{\mu\nu}(q)\Delta(q^2)+\xi\frac{q_\mu q_\nu}{q^4}\right]; \qquad
\Delta^{-1}_{\mu\nu}(q)=i\left[P_{\mu\nu}(q)\Delta^{-1}(q^2)+\xi q_\mu q_\nu\right]
\label{prop}
\ee
and $P_{\mu\nu}(q)=g_{\mu\nu}-q_\mu q_\nu/q^2$ 
the dimensionless transverse projector.  
The scalar co-factor $\Delta(q^2)$ is 
related to the all-order gluon self-energy $\Pi_{\mu\nu}(q)=P_{\mu\nu}(q)\Pi(q^2)$  through
\be
\Delta^{-1}({q^2})= q^2+i\Pi(q^2) \equiv q^2\bcj(q^2), 
\ee
where the quantity $\bcj(q^2)$ is defined in order to maintain a notational proximity with~\cite{Ball:1980ax}.

Note that the PT  makes no {\it ab initio}  
reference to a background gluon; at the level of the Yang-Mills Lagrangian there is only one gauge field, $A$, which is quantized in the 
usual way, by means of a linear gauge-fixing term of the type $\frac{1}{2\xi}(\partial_{\mu} A^{\mu})^2$ (the $R_{\xi}$ gauges). 
However, the decomposition~(\ref{deco}) assigns 
right from the start a special role to the leg carrying the momentum $q$, that is to be eventually identified with the background leg.  
Thus, unlike $\gtree_{\alpha\mu\nu}(q,r,p)$, which is Bose-symmetric with respect to all its three legs,  
the vertex $\gtreeb$ is in fact Bose-symmetric {\it only} with respect to the (quantum) $\mu$ and $\nu$ legs. In addition, 
it satisfies the simple Ward identity
\be
iq^\alpha \gtreeb_{\alpha\mu\nu}(q,r,p)=\Delta^{-1}_{0\,\,\mu\nu}(p)- \Delta^{-1}_{0\,\,\mu\nu}(r) \,,
\ee
where $\Delta_{0}^{\mu\nu}(q)$ is the tree-level version of the $\Delta^{\mu\nu}(q)$ given in Eq.~(\ref{prop}).
In higher orders, the $BQQ$ vertex is constructed  
through the systematic triggering of internal STIs in the diagrams of the conventional 
(higher order) three-gluon vertex (see the tree last items in ~\cite{Cornwall:1981zr}). 


On the other hand, when quantizing the theory within the BFM~\cite{Abbott:1980hw}, the 
$BQQ$ vertex $\Gamma_{\widehat{A}AA}$ arises directly, as a consequence 
of the splitting of the classical gauge field into a background 
and a quantum component, $A\to A+\widehat{A}$. 
In addition, one introduces a 
special gauge-fixing function that is linear in the quantum field $A$, and  
preserves gauge invariance with respect to the background field $\widehat{A}$
(the corresponding gauge-fixing parameter is denoted $\xi_Q$). 
Let us denote the {\it full} 
$BQQ$ vertex by $\Gamma_{\widehat{A}^a_\alpha A^b_\mu A^c_\nu}(q,r,p)$, and factor out the usual 
coupling and color structure,    
\be
i\Gamma_{\widehat{A}^a_\alpha A^b_\mu A^c_\nu}(q,r,p) = gf^{abc} \gfullb_{\alpha\mu\nu}(q,r,p)\,.
\ee 
At tree-level, $\gfullb_{\alpha\mu\nu} \to \gtreeb_{\alpha\mu\nu}(\xi\to \xi_Q )$, i.e. it is given by 
the expression for $\gtreeb_{\alpha\mu\nu}$ in Eq.~(\ref{deco}), after the simple replacement $\xi \to \xi_Q$.

In order to cast the upcoming WI and STIs into a more compact form, 
it is convenient to consider instead of $\gfullb_{\alpha\mu\nu}(q,r,p)$ the minimally modified vertex 
\be
\bqq_{\alpha\mu\nu}(q,r,p)= \gfullb_{\alpha\mu\nu}(q,r,p) + (1/\xi_Q)\Gamma^{{\rm P}}_{\alpha\mu\nu}(q,r,p)\,.
\ee
Evidently, $\bqq_{\alpha\mu\nu}(q,r,p)$ and $\gfullb_{\alpha\mu\nu}(q,r,p)$ differ only at tree level; 
specifically, using  Eq.~(\ref{deco}), we see immediately that 
\be
\bqq_{\alpha\mu\nu}^{(0)}(q,r,p)^{(0)}  = \gtree_{\alpha\mu\nu}(q,r,p).  
\ee
Incidentally, notice that $\bqq_{\alpha\mu\nu}(q,r,p)$ coincides with the vertex  
appearing in the SDE for the gluon propagator, when projecting to the Landau gauge~\cite{Aguilar:2008xm}.

\begin{figure}[!t]
\begin{center}
\includegraphics[scale=.7]{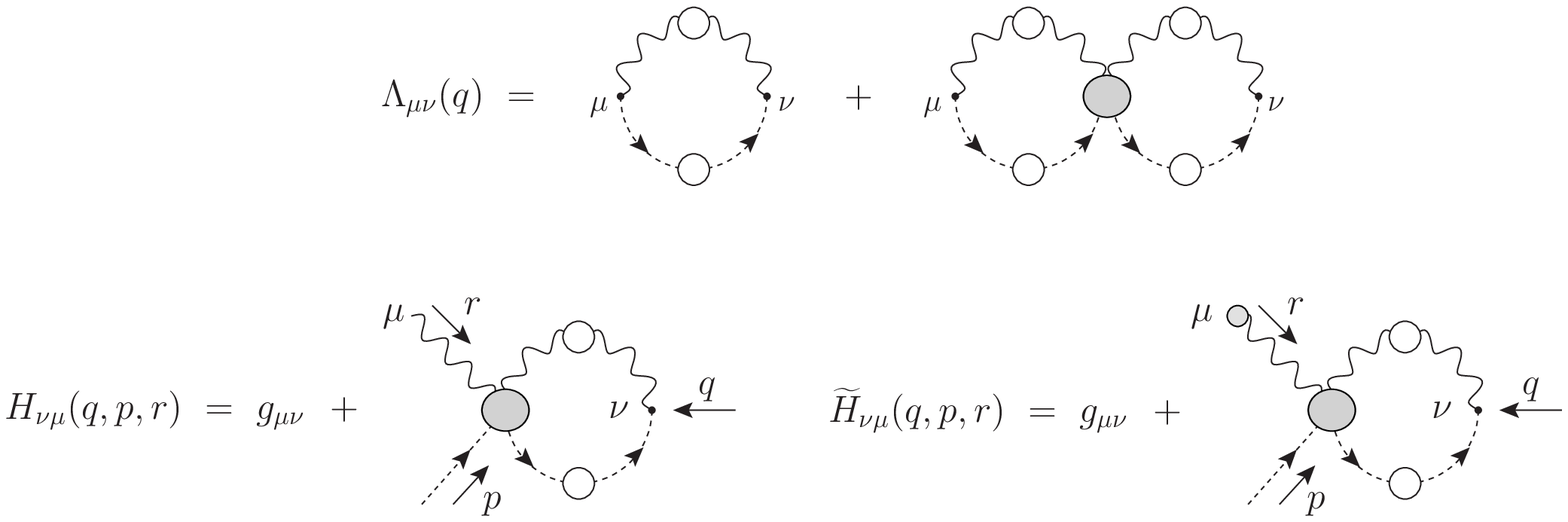}
\end{center}
\caption{\label{H-Htilde}Definitions and conventions of the auxiliary functions $\Lambda$, 
$H$ and $\widetilde{H}$. The color and coupling dependence for 
the combination shown, $c^a(q)A_\mu^b(r)A_\nu^{*c}(p)$, is $gf^{acb}$. 
White blobs represent connected Green's functions, while gray blobs denote 
the two-gluon--two-ghost kernel. Note that the kernel  
is one-particle irreducible with respect to perpendicular cuts.
}
\end{figure}

Thus, $\bqq$ 
satisfies a (ghost-free) WI when contracted with the momentum $q_\alpha$ of the 
background gluon, while it satisfies a STI  when contracted with 
the momentum of the quantum gluons ($r_\mu$ or $p_\nu$). They are given by (second item in~\cite{Aguilar:2006gr}) 
\bea
q^\alpha\bqq_{\alpha\mu\nu}(q,r,p)&=&p^2\bcj(p^2)P_{\mu\nu}(p)-r^2\bcj(r^2)P_{\mu\nu}(r)
\nonumber \\
r^\mu\bqq_{\alpha \mu \nu}(q,r,p)&=&F(r^2)\left[q^2\bcjb(q^2)P_\alpha^\mu(q)H_{\mu\nu}(q,r,p)-
p^2\bcj(p^2)P_\nu^\mu(p)\widetilde{H}_{\mu\alpha}(p,r,q)\right] \nonumber \\
p^\nu\bqq_{\alpha \mu \nu}(q,r,p)&=&F(p^2)\left[r^2\bcj(r^2)P_\mu^\nu(r)\widetilde{H}_{\nu\alpha}(r,p,q)-
q^2\bcjb(q^2) P_\alpha^\nu(q)H_{\nu\mu}(q,p,r)\right].
\label{STIs}
\eea
In the above equations, $F(q^2)$ is the ghost dressing function, related to the ghost propagator $D(q^2)$ through
\be
i D^{ab}(q^2)= i \delta^{ab}\frac{F(q^2)}{q^2},
\ee
while the propagator $\widetilde{\Delta}$ is related to the conventional one, $\Delta(q^2)$, through the so-called 
``background quantum identity''~\cite{Grassi:1999tp} 
\be
\Delta(q^2)=\left[1+G(q^2)\right]\widetilde\Delta(q^2).
\ee
The function $G$ appearing above is the $g_{\mu\nu}$ co-factor in the Lorentz decomposition of 
the auxiliary function $\Lambda_{\mu\nu}(q)$ defined through  
\bea
\Lambda_{\mu\nu}(q)&=&\int_k\!\Delta_\mu^\sigma(k)D(q-k)H_{\nu\sigma}(-q,q-k,k)\nonumber\\
&=&g_{\mu\nu}G(q^2)+\frac{q_\mu q_\nu}{q^2}L(q^2),
\eea
and shown in Fig.~\ref{H-Htilde}, together with the definitions and conventions of the functions $H$ and $\widetilde{H}$.
Therefore, requiring the vertex Ansatz to satisfy the STIs above implies that in its expression certain combinations of 
the ghost auxiliary functions $G$, $H$ and $\widetilde{H}$ will also appear.


\section{Identities of the ghost sector}

In this section we explain the field-theoretic origin of a set of constraints 
whose validity must be invoked 
when attempting  to solve the WI and STI of  Eq.~(\ref{STIs}), in order to construct the 
$BQQ$ vertex. As we will shortly, the underlying reason for having to resort to these 
constraints is the fact that the resulting system has more equations than unknowns, a fact known  
from the early work of Ball and Chiu~\cite{Ball:1980ax} on the conventional three-gluon vertex.

To begin with, we observe that, since both the $R_\xi$ and the BFM are linear gauge fixing conditions, 
there exists  a constraint coming from the equation of motion of 
the Nakanishi-Lautrup fields (the so called ghost or Faddeev-Popov equation),
 which, in turn, implies that the functions $H$ and $\widetilde{H}$ are 
related to the corresponding gluon-ghost 
trilinear vertices $\Gamma_\alpha$ and $\widetilde{\Gamma}_\alpha$. In particular one has (second item in~\cite{Aguilar:2006gr})
\bea
p^\nu H_{\nu\alpha}(p,r,q)+\Gamma_{\alpha}(r,q,p)&=&0 \nonumber \\
p^\nu \widetilde{H}_{\nu\alpha}(p,r,q)+\widetilde{\Gamma}_{\alpha}(r,q,p)&=&r_\alpha F^{-1}(r^2), 
\eea
where at tree-level
\be
 \Gamma^{(0)}_{\alpha}(r,q,p)=-p_\alpha; \qquad
 \widetilde{\Gamma}^{(0)}_{\alpha}(r,q,p)=(r-p)_\alpha.
\ee
 
In addition, while the function $\widetilde{H}$ satisfies the WI
\be
q^{\alpha}\widetilde{H}_{\nu\alpha}(p,r,q)=-p_\nu F^{-1}(p^2)-r_\nu F^{-1}(r^2),
\label{HWI}
\ee
the function $H$ fulfills  the STI 
\be
q^{\alpha}H_{\nu\alpha}(p,r,q)=-F^{-1}(q^2)\left[p_\nu F^{-1}(p^2) C(q,r,p)+r^\alpha F^{-1}(r^2)H_{\nu\alpha}(p,q,r)\right],
\label{HSTI}
\ee
where $C$ represents yet another ghost auxiliary function $\Gamma_{ccc^*}$ involving two ghosts and an anti-ghost field (with momentum $p$).
To proceed further,  we decompose  $H$ and $\widetilde{H}$ in terms of their basic tensor forms  
\bea
H_{\nu\alpha}(p,r,q)&=&g_{\alpha\nu}a_{qrp}-r_\alpha q_\nu b_{qrp}+q_\alpha p_\nu c_{qrp}+q_\nu p_\alpha d_{qrp}+p_\alpha
p_\nu e_{qrp}, \nonumber \\
\widetilde{H}_{\nu\alpha}(p,r,q)&=&g_{\alpha\nu}\widetilde{a}_{qrp}-r_\alpha q_\nu \widetilde{b}_{qrp}+q_\alpha p_\nu \widetilde{c}_{qrp}+q_\nu p_\alpha \widetilde{d}_{qrp}+p_\alpha p_\nu \widetilde{e}_{qrp}.
\label{Hdec}
\eea
where, following the notation of~\cite{Ball:1980ax} we have introduced the short-hand $a_{qrp}$ for $a(q,r,p)$, and similarly for all other form factors appearing in (\ref{Hdec}). Then, one can use the identities (\ref{HWI}) and (\ref{HSTI}) in order to constrain certain combinations of these form factors. Indeed, from the WI~(\ref{HWI}) one finds
\bea
\widetilde{a}_{qrp}-(q\cdot r)\widetilde{b}_{qrp}+(q\cdot p)\widetilde{d}_{qrp}&=&F^{-1}(r^2)\nonumber \\
q^2\widetilde{c}_{qrp}+(q\cdot p)\widetilde{e}_{qrp}+F^{-1}(p^2)&=&F^{-1}(r^2),
\label{Hhatconstr}
\eea
while the STI~(\ref{HSTI}) gives
\bea
F(r^2)\left[a_{qrp}-(q\cdot r)b_{qrp}+(q\cdot p)d_{qrp}\right]&=&F(q^2)\left[a_{rqp}-(q\cdot r)b_{rqp}+(p\cdot r)d_{rqp}\right] \nonumber \\
F^{-1}(q^2)\left[q^2c_{qrp}+(q\cdot p) e_{qrp}\right]+F^{-1}(p^2)C_{prq}&=&F^{-1}(r^2)\left[a_{rqp}-(q\cdot r)b_{rqp}  +(p\cdot r)d_{rqp} \right]\nonumber \\
&-&F^{-1}(r^2)\left[r^2 c_{rqp}+ (p\cdot r)e_{rqp}\right],
\label{Hconstr}
\eea
where, according to our conventions, $C_{prq}=C(q,r,p)$.

Incidentally, notice that the first equation in~(\ref{Hconstr}), together with its cyclic permutations 
of momenta and indices, represent the three constraints found in~\cite{Ball:1980ax} [{\it viz.} Eq.~(2.10) in that article] 
as necessary conditions 
to determine the  conventional three-gluon vertex from solving 
the corresponding STIs (the system is in fact overconstrained displaying more equations than unknowns); 
indeed we see from the above that  such constraints are a consequence of the STI satisfied by 
the $H$ function (in~\cite{Ball:1980ax} these constraints were explicitly verified at the one-loop level only). 

Finally, let us conclude by observing that $H$ and $\widetilde{H}$ are related by the BQI
\bea
\widetilde{H}_{\nu\alpha}(p,r,q)&=&\left[g_\alpha^\gamma+\Lambda_\alpha^\gamma(q)\right]H_{\nu\alpha}(p,r,q)-r^\gamma F^{-1}(r^2)N_{\alpha\gamma\nu}(q,r,p)\nonumber \\
&+&p_\nu F^{-1}(p)O_\alpha(q,r,p),
\eea
where the auxiliary functions $N$ and $O$ are related to certain auxiliary functions involving the background source $\Omega$, namely
$\Gamma_{\Omega AA^*}$ and $\Gamma_{\Omega c c^*}$.

\section{Solving the Ward and Slavnov-Taylor identities}

In what follows we determine an 
Ansatz for  $\widetilde{\gb}$ by solving the WI and  
STIs given in Eq.~\ref{STIs}. 
It is well-known  that the gauge technique, in general, 
can only furnish information about the longitudinal part of any vertex, leaving its  
transverse (automatically conserved) part completely undetermined.     
This fact, in turn, is known to be of limited importance in the infrared (in the presence of a mass gap!),
but is essential for the multiplicative renormalizability of the 
resulting SDEs. For the purposes of this work, we will ignore such refinements, 
settling for subtractive renormalizability only.  

Therefore, following~\cite{Ball:1980ax}, the 
14 possible tensorial structures necessary for describing a general 
three-gluon vertex are separated into two groups,  
10 of them spanning the longitudinal part of the vertex, and the remaining 4 the (totally) transverse part;  
then only the former group is considered.

Specifically, in the basis of~\cite{Ball:1980ax}
the longitudinal part of $\bqq_{\alpha\mu\nu}(q,r,p)$ has the form 
\be
\bqq^{(\ell)}_{\alpha\mu\nu}(q,r,p)=\sum_{i=1}^{10}X_i(q,r,p)\ell^i_{\alpha\mu\nu}(q,r,p),
\label{tenlon}
\ee
with the explicit form of the tensors $\ell^i$ given by
\be
\begin{tabular}{lll}
$\ell^1_{\alpha\mu\nu} =  (q-r)_{\nu} g_{\alpha\mu}$
& 
$\ell^2_{\alpha\mu\nu} =  - p_{\nu} g_{\alpha\mu}$\hspace{.75cm}
&
$\ell^3_{\alpha\mu\nu} =  (q-r)_{\nu}[q_{\mu} r_{\alpha} -  (q\cdot r) g_{\alpha\mu}] $\\
$\ell^4_{\alpha\mu\nu} = (r-p)_{\alpha} g_{\mu\nu}$
&
$\ell^5_{\alpha\mu\nu} =  - q_{\alpha} g_{\mu\nu}$
&
$\ell^6_{\alpha\mu\nu} =  (r-p)_{\alpha}[r_{\nu} p_{\mu} -  (r\cdot p) g_{\mu\nu}]$
\\
$\ell^7_{\alpha\mu\nu} =  (p-q)_{\mu} g_{\alpha\nu}$
&
$\ell^8_{\alpha\mu\nu} = - r_{\mu} g_{\alpha\nu}$
&
$\ell^9_{\alpha\mu\nu} = (p-q)_{\mu}[p_{\alpha} q_{\nu} -  (p\cdot q) g_{\alpha\nu}]$
\\
 $\ell^{10}_{\alpha\mu\nu} = q_{\nu}r_{\alpha}p_{\mu} + q_{\mu}r_{\nu}p_{\alpha}$. & &
\end{tabular}
\label{Ls}
\ee
Notice that excluding $\ell^{10}$, each of the remaining $\ell^{i+3}$ can be obtained by the corresponding $\ell^i$ through cyclic permutation of momenta and indices; in addition,  Bose symmetry with respect to the quantum legs requires that $\Gamma$ change sign under the interchange of  the corresponding Lorentz indices and  momenta, thus implying the relations
\be
\begin{tabular}{lll}
$X_1 (q,p,r) =  X_7 (q,r,p)$\hspace{0.75cm}
&
$X_2 (q,p,r) =   - X_8 (q,r,p)$\hspace{0.75cm}
&
$X_3 (q,p,r) =   X_9 (q,r,p)$
\\
$X_4 (q,p,r) =  X_4 (q,r,p)$
&
 $X_5 (q,p,r) =  - X_5 (q,r,p)$
&
$X_6 (q,p,r) =   X_6 (q,r,p)$
\\
$X_{10} (q,p,r) =   - X_{10} (q,r,p).$ & &
\end{tabular}
\label{boserel}
\ee

The form factors $X_i$ are then fully determined 
by solving the system of linear equations generated by the identities given in Eq.~(\ref{STIs}). 
The procedure is conceptually straightforward, but operationally rather cumbersome. 
One first substitutes on the lhs of Eq.~(\ref{STIs}) the general tensorial decomposition of  
$\bqq^{(\ell)}_{\alpha\mu\nu}(q,r,p)$ given in Eq.~(\ref{tenlon}), and then equates 
the coefficients of the resulting tensorial structures to those appearing on the rhs. 
Thus, one obtains a system of equations expressing the form factors $X_i(q,p,r)$ 
in terms of combinations of quantities such as $J$, $F$, etc. 

In what follows we will only report the set of independent equations, i.e., we will omit
equations that can be obtained from existing ones by implementing the change $p \leftrightarrow r$ 
and using the constraints of ~(\ref{boserel}). Thus, for example, the equation 
$X_7 + X_8 + (q\cdot r)X_{10} =  J(p)$ does not form part of the set of independent equations, 
because it can be obtained from the second equation in Eq.(\ref{absys}) below, by carrying out the 
aforementioned transformation, and using the corresponding relations from Eq.~(\ref{boserel}).

Thus, from the abelian WI one obtains the following 4 equations  
\bea
(p^2-r^2) X_4  - q^2 X_5 - (r\cdot p) (p^2-r^2)X_6 &=&  p^2J(p) - r^2J(r)
\nonumber\\
X_1 - X_2 - (q\cdot p)X_{10} &=&   J(r)
\nonumber\\
X_1 + X_2 - X_7 + X_8 &=&  0
\nonumber\\
2 X_1 + (p^2-r^2) X_6 - 2 X_7  + q^2 X_{10} &=&  0,
\label{absys}
\eea
where the form of the second equation has been simplified by making use of the third. 

Similarly, from the non-abelian STI one obtains 5 equations, namely 
\bea
(r^2-q^2) X_1  - p^2 X_2 - (q\cdot r) (r^2-q^2)X_3 &=&  F(p)
\left[{\widetilde a}_{qpr} r^2 J(r) - a_{rpq} q^2{\widetilde J}(q)\right]
\nonumber\\
(r^2-q^2) X_3 - 2  X_4 + 2 X_7 + p^2 X_{10} &=& F(p)
\left[({\widetilde b}_{qpr}+ {\widetilde d}_{qpr}) r^2 J(r) 
- (b_{rpq}+d_{rpq}) q^2{\widetilde J}(q)\right]
\nonumber\\
- X_7 +  X_8 + (r\cdot p) X_{10} &=&   F(p)
\left\{(r\cdot p) {\widetilde b}_{qpr} J(r) 
- \left[a_{rpq} +(q\cdot r){d}_{rpq}\right]{\widetilde J}(q)
\right\}
\nonumber\\
X_4 + X_5 + (q\cdot p)X_{10} &=& F(p)
\left\{\left[{\widetilde a}_{qpr} +(q\cdot r) {\widetilde d}_{qpr}\right] J(r) 
- (q\cdot p) b_{rpq} {\widetilde J}(q)\right\}
\nonumber\\
-X_4 + X_5 + X_7 + X_8  &=&  (q\cdot r)F(p)
\left[-{\widetilde b}_{qpr} J(r) 
+ b_{rpq} {\widetilde J}(q)\right].
\label{nabsys}
\eea
Clearly, there are 5 additional equations, obtained from the second STI; however, they too  
can be obtained from the set of equations~(\ref{nabsys})
by imposing the transformation $r \leftrightarrow p$ and using the 
relations given in Eq.~(\ref{boserel}), and are therefore omitted. 

As anticipated, we have more equations than form factors [remember the constraints of Eq.~(\ref{boserel})!], 
and therefore the appearance  of a set of non-trivial constraints
for the ghost sector. It turns out that these constraints are {\it precisely} those furnished by 
Eq.~(\ref{Hhatconstr}) and the 
first relation of Eq.~(\ref{Hconstr}). 
Therefore the system can be solved, and one finds a solution of the type presented in~\cite{Ball:1980ax} 
with a modified ghost-sector, reading
\bea
X_1(q,r,p) &=&  \frac{1}{4}{\widetilde J}(q) \left\{ 
- p^2 b_{prq} F(r) + 
[2 a_{rpq} + p^2 b_{rpq} + 2 (q\cdot r) d_{rpq} ]F(p) \right\}    
\nonumber\\
&+& \frac{1}{4} J(r)\left[ 2 +  (r^2-q^2) {\widetilde b}_{qpr} F(p)\right]
 + \frac{1}{4} J(p)\, p^2 \,{\widetilde b}_{qrp} F(r) 
\nonumber\\
X_2(q,r,p)  &=&  \frac{1}{4}{\widetilde J}(q)\left\{
(q^2- r^2) b_{prq} F(r) + 
[2 a_{rpq} + (r^2-q^2)b_{rpq} + 2 (q\cdot r) d_{rpq} ] F(p)\right\}  
\nonumber\\
&+& \frac{1}{4} J(r)\left[ - 2 +  p^2 {\widetilde b}_{qpr} F(p)\right]
 + \frac{1}{4} J(p)\, (r^2- q^2)\,{\widetilde b}_{qrp} F(r) 
\nonumber\\
X_3(q,r,p) &=& \frac{F(p)}{q^2-r^2} 
\left\{{\widetilde J}(q)\left[a_{rpq} - (q\cdot p) d_{rpq} \right] 
-  J(r) \left[{\widetilde a}_{qpr} - (r\cdot p){\widetilde d}_{qpr}\right] \right\}
\nonumber\\
X_4(q,r,p) &=& \frac{1}{4}{\widetilde J}(q) q^2 \left[b_{prq} F(r) + b_{rpq} F(p)\right]
+ \frac{1}{4} J(r)  \left[2 - q^2 {\widetilde b}_{qpr} F(p)\right]
\nonumber\\
&+& \frac{1}{4} J(p) \left[2 - q^2 {\widetilde b}_{qrp} F(r)\right]
\nonumber\\
X_5(q,r,p) &=& \frac{1}{4}{\widetilde J}(q) (p^2-r^2) \left[b_{prq} F(r) + b_{rpq} F(p)\right]
+ \frac{1}{4} J(r)  \left[2 +(r^2-p^2) {\widetilde b}_{qpr} F(p)\right]
\nonumber\\
&-& \frac{1}{4} J(p) \left[2 +(p^2-r^2) {\widetilde b}_{qrp} F(r)\right]
\nonumber\\
X_6(q,r,p) &=& \frac{J(r)-J(p)}{r^2-p^2}
\nonumber\\
X_7(q,r,p) &=& X_1(q,p,r)
\nonumber\\
X_8(q,r,p) &=& - X_2(q,p,r)
\nonumber\\
X_9(q,r,p) &=& X_3(q,p,r)
\nonumber\\
X_{10}(q,r,p) &=& \frac{1}{2}\left\{ 
{\widetilde J}(q) \left[b_{prq} F(r) - b_{rpq} F(p)\right]
+   J(r) F(p) {\widetilde b}_{qpr} - J(p) F(r) {\widetilde b}_{qrp} \right\}.
\label{X10}
\eea

\section{\label{concl}Conclusions}

We have presented a complete Ansatz for the $BQQ$ three-gluon vertex,  
which is in absolute conformity with both the WI and the STI given in Eq.~(\ref{STIs}). 
An important step in this construction is the formal, all-order derivation 
of the crucial constraints relating the various form factors of the ghost Green's function, 
involved in the STI of Eq.~(\ref{STIs}), an indispensable step for realizing this construction.

It is important to emphasize that the Ansatz for the $BQQ$ vertex presented here 
is valid for any value of the 
gauge-fixing parameter used to quantize the theory. Indeed, even though 
the various ingredients appearing in the solution Eq.~(\ref{X10}), such as $J$, $F$, etc, 
 depend explicitly on $\xi$ (or on $\xi_Q$), the precise functional dependence of the form factors 
$X_{i}$ on $J$, $F$, etc, given in  Eq.~(\ref{X10}), is valid for any $\xi$, given that it originates 
from the solution of the WI and STI of Eq.~(\ref{STIs}), whose form is  $\xi$-independent. 

This last point is 
particularly important given the existing perspectives~\cite{Cucchieri:2009kk} of carrying out large-volume
lattice simulations in covariant gauges other than the Landau gauge ($\xi\neq 0$). In particular, 
the possibility of simulating propagators in the background gauges 
(especially the background Feynman gauge, $\xi_Q=1$) 
opens up the exiting possibility of studying central objects of the PT 
on the lattice~\cite{Cucchieri:2011pp}.

An additional important point, not addressed here, is related to the way the $BQQ$ vertex 
triggers the Schwinger mechanism~\cite{Schwinger:1962tn}, 
which, in turn, is responsible for the dynamical generation 
of a gluon mass. As is well-known~\cite{Jackiw:1973tr}, the relevant three-gluon vertex (in this case the  $BQQ$ vertex )
must contain longitudinally coupled massless poles, in order for gauge invariance to be preserved. 
We emphasize that the Ansatz presented here does not incorporate such poles, which must be supplied 
at a subsequent step. We hope to accomplish this task in the near future.

{\it Acknowledgments:} 

I thank the organizers of this Workshop for their warm hospitality 
and stimulating environment. 
This research was supported by the European FEDER and  Spanish MICINN under grant FPA2008-02878,
and the Fundaci\'on General of the UV.

\end{document}